# Observation of stacked vortex structure with number controllable Majorana zero mode in $Bi_2Te_3$/FeTe heterojunction

Xinru Han[1, 2, *], Zhaokun Luo[1, *], Quanxin Hu[7, 1, *], Xingjian Zhu[1], Xiaohong Pan[8], Dan Mu[1, 9], Xinyu Liang[1], Zulin Xu[1], Qiheng Wang[1], Ryan L. Kong[1], Mingyuan Li[1], Zhengliang Zhang[1], Fazhi Yang[12], Mojun Pan[1], Xin Liu[1, 3, 4, 6], Qiangqiang Gu[1], Baiqing Lv[1, 4, 5, †], Gan Wang[2, 10, †], Hong Ding[1, 3, 11, †]

*[1] Tsung-Dao Lee Institute, School of Physics and Astronomy, and State Key Laboratory of Micro-nano Engineering Science, Shanghai Jiao Tong University, Shanghai 201210, China*

*[2] Department of physics, Southern University of Science and Technology, Shenzhen 518055, China*

*[3] Hefei National Laboratory, Hefei 230088, China*

*[4] Shanghai Research Center for Quantum Sciences, Shanghai 201315, China*

*[5] Zhangjiang Institute for Advanced Study, Shanghai Jiao Tong University, Shanghai 200240, China*

*[6] School of Physics and Wuhan National High Magnetic Field Center, Huazhong University of Science and Technology, Wuhan, Hubei 430074, China*

*[7] Department of Physics, College of Physical Science and Technology, Xiamen University, Xiamen 361005, China*

*[8] College of Physics & Optoelectronic Engineering, Department of Physics, Jinan University, Guangzhou 510632, China.*

*[9] Institute for Quantum Science and Technology, Shanghai University, Shanghai 200444, China*

*[10] State Key Laboratory of Quantum Functional Materials, Southern University of Science and Technology, Shenzhen 518055, China*

*[11] New Cornerstone Science Laboratory, Shanghai 201210, China*

*[12] Department of Physics, City University of Hong Kong; Kowloon, Hong Kong, China*

[*] These authors contributed equally to this work.

[†] Corresponding authors: baiqing@sjtu.edu.cn; wangg@sustech.edu.cn; dingh@sjtu.edu.cn

**Abstract:**

Creation and control of Majorana zero modes (MZMs) in solid devices is the key challenge in realizing fault-tolerant topological quantum computation. Here we report spectroscopic evidence for number controllable MZMs in stoichiometric one quintuple layer $Bi_2Te_3$/FeTe heterostructures, together with the observation of stacked vortices in superconducting islands. Low-temperature scanning tunnelling microscopy and spectroscopy reveal a long-period moiré superlattice arising from lattice mismatch, which induces nanoscale modulation of superconducting gaps. Magnetic vortices exhibit a stacked-core structure with two layer-derived bound-state branches: a dispersive branch evolving into an X-shaped pattern away from the vortex center, and a quantum-limit branch characterized by a spatially unsplit MZMs together with discrete integer-quantized Caroli–de Gennes–Matricon states. Notably, MZMs are observed in all measured vortices, and their number can be tuned by magnetic field and the size of $Bi_2Te_3$ islands. These results establish stoichiometric 1QL-$Bi_2Te_3$/FeTe as a patterned superconducting interface that enables isolated and few-vortex Majorana configurations, providing a controlled platform for fuse MZMs.

**Main text:**

Majorana zero modes (MZMs) are proposed as building blocks for fault-tolerant topological quantum computation owing to their non-Abelian exchange statistics(*1–3*). A practical MZM platform is therefore expected to meet several stringent criteria, including a large topological gap protecting the zero-energy modes from quasiparticle excitations, a high probability of MZM formation, and precise control over their number and spatial distribution(*4–6*). A range of material systems have been proposed to host MZMs, including intrinsic *p*-wave superconductors(*7–10*), heterostructures combining strong spin orbit coupling (SOC) semiconductor nanowire with *s*-wave superconductor (SC)(*11*, *12*), topological insulator/superconductor heterostructures(*7*, *13–15*), and iron-based superconductors(*16–19*). In many of these platform for realizing MZM, strong SOC semiconductor nanowire/s-wave superconductor have been theoretically predicted to host a pair of MZMs localized at the two ends of the wire. Owing to their high compatibility with modern semiconductor growth and nanofabrication technologies(*20*, *21*), these platforms have witnessed rapid progress in recent years. However, existing nanowire-based platforms remain limited by several major limitations, including a small topological protection gap and a low superconducting transition temperature, while conclusive evidence for the existence of MZMs in these systems is still lacking(*22*, *23*). Iron-based superconductors offer an alternative route due to their comparatively large superconducting gaps, which can provide stronger topological protection. However, they are typically constrained by materials specific limitations. The archetypal Fe(Te, Se) system suffers from alloy-induced chemical inhomogeneity, leading to non-uniform spatial distributions of MZMs(*24*, *25*). Other stoichiometric candidates, such as $CaKFe_4As_4$, are limited by the polar surface problem(*26*). Therefore, advancing from isolated spectroscopic signatures to braidable topological qubits will likely require a stoichiometric, structurally well-defined superconducting platform that simultaneously enables robust topological protection and deterministic control of MZMs.

Heterostructures based on iron-based superconductors may provide a key route toward overcoming these limitations. However, constructing heterostructures in which iron-based superconductors serve as the superconducting substrate inevitably leads to vortex formation throughout the whole heterostructure under the applied magnetic field. This results in dense and uncontrollable Abrikosov vortex configurations, making it difficult to isolate and deterministically manipulate a controllable number of vortices required for studying Majorana interactions and fusion processes(*6*, *27*, *28*). This limitation indicates that an ideal platform should confine superconductivity to spatially engineered interfacial regions, ensuring that vortex degrees of freedom are restricted to these islands where their number and arrangement can be controlled. In this context, the heterostructures with interfacial superconductivity emerging from non-superconducting constituents have become a promising alternative(*29*, *30*). Experimentally, interfacial superconductivity has been established in several iron chalcogenide/topological insulator heterostructures, most notably $Sb_2Te_3$/FeTe and $Bi_2Te_3$/FeTe systems(*31–33*). In these materials, neither constituent is superconducting; nevertheless, superconductivity emerges at the interface, providing a possible promising route toward topological superconducting phases.

Here we realize a stoichiometric moiré superconducting interface by growing high-quality one quintuple layer (1QL) $Bi_2Te_3$/FeTe heterostructure using molecular beam epitaxy, and probe its electronic structure using low-temperature scanning tunnelling microscopy and spectroscopy (STM/S). The heterostructure exhibits a long-period moiré superlattice arising from the structural mismatch between the hexagonal Te network of $Bi_2Te_3$ and the square Te lattice of FeTe. This moiré potential imprints a periodic modulation on two distinct superconducting gaps. Under the perpendicular magnetic field, spectroscopic imaging reveals vortices with a stacked-core structure, manifested by two coexisting branches of vortex bound states that originate respectively from the $Bi_2Te_3$ and FeTe layers. The $Bi_2Te_3$ derived branch disperses away from the vortex center into an X-shaped pattern, whereas the FeTe derived branch remains in the quantum limit and exhibits a spatially unsplit zero-bias conductance peak accompanied by discrete integer-quantized Caroli–de Gennes–Matricon (CdGM) states. These spectroscopic signatures are consistent with Majorana zero modes(*34*, *35*). By varying the magnetic field and the lateral size of $Bi_2Te_3$ islands, we further realize controllable configurations containing one, two and three Majorana vortices. Our results establish stoichiometric 1QL-$Bi_2Te_3$/FeTe as a clean and spatially controllable superconducting interface, enabling the controlled creation of isolated and few-vortex Majorana configurations for exploring Majorana interactions and fusions.

**Interfacial superconductivity emerged in 1QL-$Bi_2Te_3$/FeTe**

The heterostructure comprises a 20 unit cell FeTe layer grown first on the bottom, followed by a 1QL-$Bi_2Te_3$ layer on top. In this heterojunction, neither FeTe nor $Bi_2Te_3$ is superconducting on its own. Figs. 1A-C schematically illustrates the 1QL-$Bi_2Te_3$/FeTe structure (see details in Methods). Although $Bi_2Te_3$ and FeTe have different lattice symmetries, hexagonal and tetragonal, respectively, their long-period lattice matching enables hybrid van der Waals epitaxy and gives rise to a $7 \times \sqrt{3}$ moiré pattern (Fig. S2). STM topography reveals atomically flat $Bi_2Te_3$ islands extending along the FeTe step terraces (Fig. S3A), indicative of step flow epitaxy, which is important for achieving homogeneous growth and reliable thickness control. A zoom in topographic image further resolves the Te-terminated $Bi_2Te_3$ surface and the corresponding $7 \times \sqrt{3}$ moiré superstructure (Fig. 1D). The Fourier transforms exhibit additional reciprocal-lattice peaks associated with this moiré periodicity (Fig. 1E).

Superconductivity is confined to the $Bi_2Te_3$ covered heterostructure regions and is absent in the uncovered areas. By our line profile cross the edge between $Bi_2Te_3$ island and bare FeTe surface, sharp superconductivity transition can be observed (Figs. S3E-F). Representative *dI/dV* spectra acquired over a narrow bias range (±10 mV) on 1QL-$Bi_2Te_3$ and bare FeTe are shown in Fig. S3D. In contrast to pristine $Bi_2Te_3$, the typical *dI/dV* spectra acquired on 1QL-$Bi_2Te_3$ surface (Fig. S4A), shows an obviously "peak-dip-hump" and "double-peak" feature. By multigap Dynes function fitting shown in Fig. S3B, we found that the multi-superconducting coherent peak is located at $\pm 1.85$ meV and $\pm 2.92$ meV, and the symmetric dip feature is located at about $\pm 4.4$ meV. Similar peak–dip–hump structures have been reported in other superconducting heterostructures, which is likely associated with the superconducting proximity effect(*36*, *37*). The spectrum is markedly different from that on bare FeTe (Fig. S3D), which shows absence of superconducting coherence peaks with a broad depression with a relatively flat bottom. Importantly, the long period $7 \times \sqrt{3}$ moiré superstructure inevitably introduces a same moiré potential in this system, and thus may add modulations to this interfacial superconductivity(*38–40*). Indeed, the line profile along the long moiré direction shown in Fig. 1H confirm this hypothesis, the superconducting gap size and the amplitude of coherent peaks along the moiré direction exhibit clear modulations with same period of moiré pattern (Fig. 1I), consistent with previous work(*41*).

**Anisotropic Abrikosov vortex with local density of states modulations**

Given the emergence of superconductivity and the proposed band topology of FeTe, magnetic vortices in 1QL-$Bi_2Te_3$/FeTe may host MZMs. To minimize possible defect-related effects, we performed the measurements in a defect-free region at 0.4 K under a perpendicular magnetic field of 1.0 T. A well-defined vortex is observed 1QL-$Bi_2Te_3$/FeTe and exhibits a pronounced anisotropic star-shaped profile (Fig. 2A), with the spatial local density of states modulations extending over more than ~40 nm along six directions. Similar star-shaped vortices have previously been reported in 1QL-$Bi_2Te_3$/Fe (Te, Se), where they were attributed to the anisotropic Fermi surface of $Bi_2Te_3$ together with vortex-vortex interactions(*42*). In our case, vortex-vortex interactions should be negligible because no second vortex appears within the superconducting $Bi_2Te_3$ island at 1.0 T. We therefore associate the star-shaped vortex primarily with the anisotropic Fermi surface of $Bi_2Te_3$(*43*), consistent with the similar structure and tunnelling conditions used here.

Whereas an anisotropic Fermi surface can naturally account for the star-shaped vortex profile, the origin of the standing wave like modulation is less clear. In this situation, the magnetic vortex acts as a strong local scatterer, generating strong scattering(*44*, *45*). Besides, inside the magnetic vortex, the interfacial superconductivity has been strongly suppressed and exhibits the single electron picture, producing possible quasiparticle interference (QPI) pattern ,

as illustrated in Fig. 2H, which is the same as the scenario of magnetic field induced pairing density wave (PDW) in cuprate vortex halo(*46*). Other alternative scenarios are discussed in the Supplementary Materials. Figs. 2A-F show local density of states (LDOS) maps and their Fourier transforms at energies both inside and outside the superconducting gap, revealing a common QPI pattern both in the superconducting and normal state energy windows under the out of plane magnetic field and showing a clear sixfold symmetry pattern, which is consistent with the QPI pattern of $Bi_2Te_3$ surface in previous work(*43*, *47*). Furthermore, statistical analysis of the scattering vectors as a function of energy (Fig. 2G) shows a same scattering vector both inside and outside the superconducting gap. Compared with the normal QPI pattern of $Bi_2Te_3$ surface (Fig. S8), the scattering vectors observed in a magnetic vortex exhibit the same magnitude as those near the Fermi surface in the original QPI pattern. These results suggest a common origin for the normal-state QPI and the vortex core LDOS modulation, consistent with the picture of magnetic field pair-breaking induced quasiparticle interference.

**Double sets of vortex bound states and robust zero bias peak**

In addition to the anisotropic vortex profile on the superconducting 1QL-$Bi_2Te_3$ surface, our high resolution spectroscopic measurements reveal a fascinating dichotomy: coexistence of dispersive and discrete vortex bound states. The first set, which resides outside the quantum limit, exhibits a continuous spatial dispersion as a function of distance from the vortex center. A higher energy feature evolves into an X-shaped dispersion away from the vortex center, reminiscent of vortex-core states observed in $NbSe_2$. The multi-peak fitting shows a linear dispersion behavior away from the vortex core (10~20 nm from the center; Fig. 3F). To clarify its origin, we fit spectra acquired near the vortex edge (Fig. S8A) using a conventional BdG framework. The extracted parameters ($k_F$~0.3 nm$^{-1}$, $E_F$~300 meV) are consistent with previous Angle-resolved photoemission spectroscopy (ARPES) results and our QPI pattern on 1QL-$Bi_2Te_3$(*48*) (Fig. S10), suggesting that the dispersing high energy states originate from the vortex in the superconducting $Bi_2Te_3$ layer. The corresponding level spacing (~0.03 meV) is far below the instrumental energy resolution (~0.37 meV at 0.4 K), consistent with the observed quasi-continuous X-shaped dispersion. This dispersive behavior is highly reminiscent of the CdGM states observed in proximity-coupled topological systems such as $Bi_2Te_3/NbSe_2$(*49*), where the large $E_F$ of the Dirac surface states result in the dispersive energy spectrum.

In stark contrast, the second set manifests as a pronounced zero bias conductance peak (ZBCP) with a full width at half maximum of about 0.5 meV and an amplitude approximately twice that of the conductance outside the gap region (Fig. 2I). Besides the ZBCP, two pairs of low-energy symmetric peaks appear near the zero energy, with an energy spacing of about 0.7 meV, suggestive of integer quantized CdGM states(*50*, *51*). Figs. 3B-E display the spatial evolution of these low energy bound states. Away from the vortex center, the intensity of the finite-energy peaks decreases and disappears at about 10 nm, whereas the ZBCP remains unsplit along different directions from the core center. To detailed explore the spatial evolution of low energy states in the vortex core, we extract five discrete states at $-2.1 \pm 0.13$ meV ($L_{-2}$), $-0.93 \pm 0.14$ meV ($L_{-1}$), $0.003 \pm 0.11$ meV ($L_0$), $0.79 \pm 0.17$ meV ($L_1$), and $1.96 \pm 0.22$ meV ($L_2$) from the spectra by multi-peak fitting, which approximately follow an integer quantized sequence (Fig. 3G). These observations provide compelling spectroscopic evidence for the presence of Majorana zero modes. We further extract the spatial profile of the ZBCP using multi-Lorentzian fitting. The decay length of the ZBCP is nearly identical along different high symmetry directions (Fig. 3H-I), in sharp contrast to the pronounced anisotropic Fermi surface of $Bi_2Te_3$ and the star-shaped morphology of the $Bi_2Te_3$ derived vortex bound states. This distinct spatial symmetry, together with the shorter characteristic decay length, indicates that the second set of vortex bound states has a different microscopic origin.

As the usual observation of MZM signatures in iron-based superconductors, a natural assignment is that this branch originates from the first FeTe layer which is also superconducting. Thus, the two classes of vortex bound states exhibit markedly different coherence lengths and spatial symmetries, consistent with their distinct layer origins, displaying a multilayer stacking vortices behavior (Fig. 4A)(*52*, *53*). Despite these differences, their vortex centers are aligned in the in plane direction, giving rise to a vertically stacked-core vortex structure across the heterostructure. In this condition, any variations of interfacial hybridization strength may modify the balance between the top layer $Bi_2Te_3$ and FeTe derived superconducting channels, resulting in a systematic evolution from dispersive high energy vortex states to localized low energy bound states. Such interface tuned spectral reconstruction suggests that interfacial coupling constitutes an effective degree of freedom for tailoring vortex electronic structure. Beyond determining the microscopic nature of vortex bound states, this interfacial engineering strategy may provide a versatile route toward controllable topological superconductivity in van der Waals heterostructures.

**Origin of Majorana zero modes**

The behavior of the second set manifests as a robust ZBCP accompanied by discrete, integer-quantized energy levels tightly localized at the vortex core is consistent with a topologically non trivial zero mode(*51*). However, the origin of this MZM is puzzling**.** The epitaxial $Bi_2Te_3$ film is strictly limited to a single quintuple layer. In the single quintuple limit, strong inter-surface quantum tunneling induces a hybridization gap, fundamentally annihilating the gapless Dirac surface states inherent to bulk $Bi_2Te_3$. Consequently, the non trivial topology requisite for the observed MZMs cannot originate from $Bi_2Te_3$.

A trivial scenario is that this MZM is generated from superconducting FeTe, like the topological superconductivity observed in $FeTe_{0.55}Se_{0.45}$, as a Dirac-like surface states has been detected by our ARPES measurements and another group. However, the topological electronic structure of $Fe_{1+y}Te$ will be strongly influenced by the amount of excess interstitial Fe. In samples with reduced interstitial Fe (Te annealed), a Dirac cone like dispersion emerges near the Fermi level (Figure S7), indicative of a non-trivial electronic structure. In contrast, this feature is absent in as-grown $Fe_{1+y}Te$ (Figure S6). This observation of Dirac like surface states upon reducing the interstitial Fe content provides further evidence for the topologically non-trivial nature of FeTe. During the growth of top layer $Bi_2Te_3$ on FeTe, the additional Te atoms will establish a Te-rich atmosphere, may produce an effect analogous to Te annealing, reducing the influence of excess Fe and restoring the non-trivial FeTe-derived band structure beneath the heterostructure, and also induces the superconductivity in FeTe reported by recent works(*55*, *56*). These observations support a scenario in which the FeTe layer underneath $Bi_2Te_3$ enter a topological superconducting phase, provides the microscopic origin of the observed Majorana zero modes.

On the other hand, the top layer $Bi_2Te_3$ may provide a charge transfer effect and enhanced SOC(*33*, *57*, *58*). These intertwined interfacial effects may substantially modify the intrinsic antiferromagnetic order and electronic correlations in FeTe, leading to pronounced band renormalization and potentially driving a topological phase transition. Moreover, the observed long-period moiré pattern is expected to generate a spatially periodic moiré potential at the $Bi_2Te_3$/FeTe interface. Such a long wavelength superlattice potential can reconstruct the low-energy electronic structure through band folding, hybridization and bandwidth renormalization, while simultaneously modifying the interfacial charge distribution, magnetic interactions and electronic correlations. These effects may provide an additional route for tuning the system across a topological phase transition and could also promote or reshape the emergent interfacial superconductivity(*38–40*, *59*). Therefore, beyond the topological superconductivity generate from iron-based superconductor family, the interface regulation also offers the possibility to realize topological superconductivity in this heterojunction.

**Summary and outlook**

In summary, our high-resolution STM spectroscopic measurements uncover a stacked-core vortex structure in stoichiometric 1QL-$Bi_2Te_3$/FeTe heterostructures, where two layer-derived branches of vortex bound states, originating from the $Bi_2Te_3$ and FeTe layers, are spatially aligned at the same in-plane vortex position. This stacked-core structure provides a setting for investigating possible interactions between distinct vortex-core states across a superconducting interface. Our results further establish a stoichiometric heterostructure route to Majorana physics, distinct from alloyed Fe(Te,Se) systems in which chemical inhomogeneity can obscure the microscopic origin and reproducibility of vortex Majorana signatures. In 1QL-$Bi_2Te_3$/FeTe, superconductivity and vortices with Majorana signatures are confined to the $Bi_2Te_3$ covered islands, enabling the vortex number to be tuned by the island size and magnetic field.

Perhaps more significantly, this natural two-dimensional Majorana platform provides the possibility to generate Majorana edge modes (MEM) at the edge of superconducting 1QL-$Bi_2Te_3$ island, and tune the presence and absence of MEM by control the vortex number in single island(*5*). This tunable MEMs satisfy the basic requirements for braiding MEM by four terminal devices (Figure S11), offer a new route to achieve topological quantum qubit(*60*, *61*). Besides, the controlled realization of isolated, paired and few-vortex configurations provides a versatile platform for studying isolated Majorana zero modes and for exploring Majorana interactions and fusion in engineered superconducting nanostructures (Figs. 4B-E).

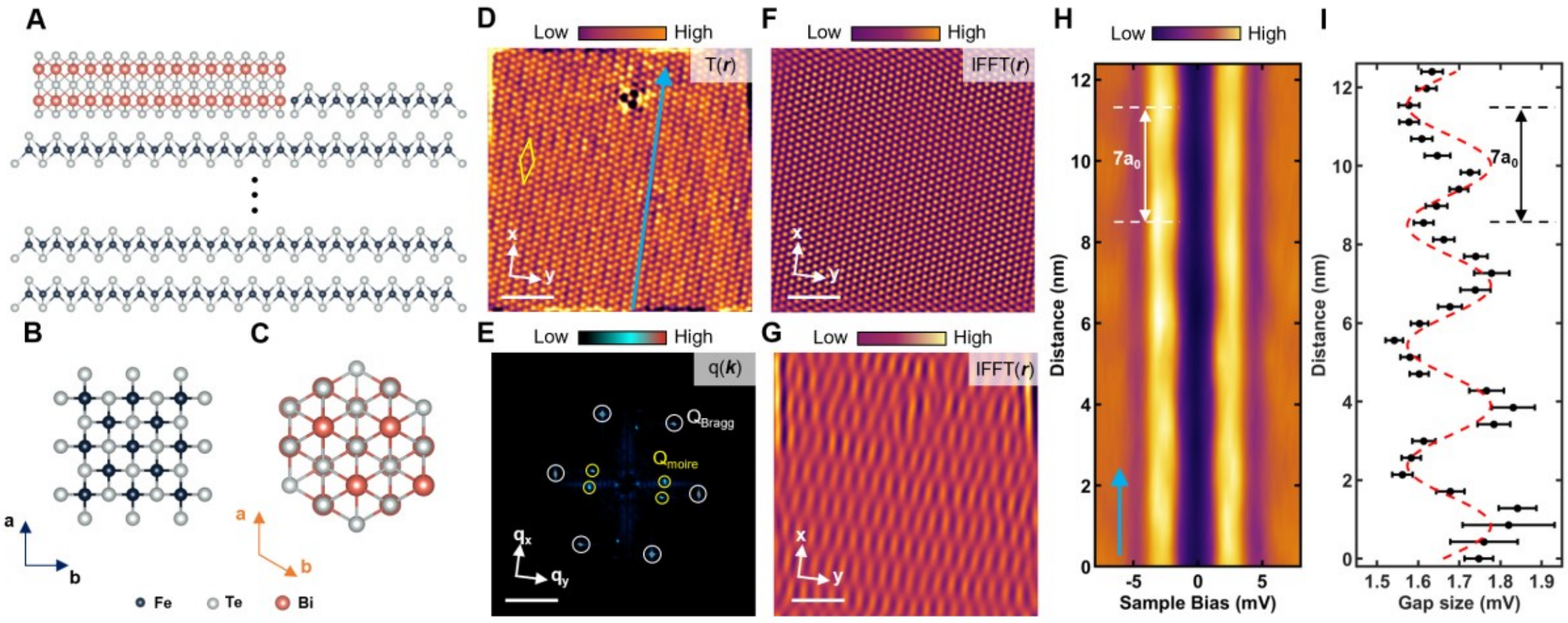

**Fig. 1| Interface superconductivity emerged in 1QL-$Bi_2Te_3$/FeTe with moiré modulation.** **A,** Side view of $Bi_2Te_3$/FeTe heterojunction. **B**-**C**, Top view of FeTe and $Bi_2Te_3$ crystal structure, FeTe have fourfold tetragonal structure while $Bi_2Te_3$ have sixfold hexagonal structure. **D,** Atomic resolution STM topography (15 nm×15 nm, $V_{bias}$ = 10 mV, $I_{Tunnel}$ = 200 pA, and T = 1.6 K) of $Bi_2Te_3$ Te-termination surface, showing moiré pattern, the yellow rhombus in (**D**) is a moiré unit cell. **E,** Fourier transform (FFT) of (**D**), six Bragg peaks of Te (white circle) and four peaks of moiré pattern (yellow circle) can be seen. **F-G,** Inversed Fourier transform (IFFT) at wavevectors of Bragg (**F**) and wavevectors of moiré pattern (**G**). **H,** Color plot of *dI/dV* spectra measured along the blue arrow in (**D**). **I**, Extracted inner superconducting gap size from *dI/dV* spectra measured along the blue arrow in (**D**), red dash line is the sine fitting of extracted gap size, showing clear modulation with 2.9 nm (about $7a_0$) moiré period. ($V_{bias}$ = 10 mV, $I_{Tunnel}$ = 200 pA, and $V_{mod}$ = 0.1 mV). Scale bar: 3 nm (**D, F, G**); 1 $Å^{-1}$ (**E**).

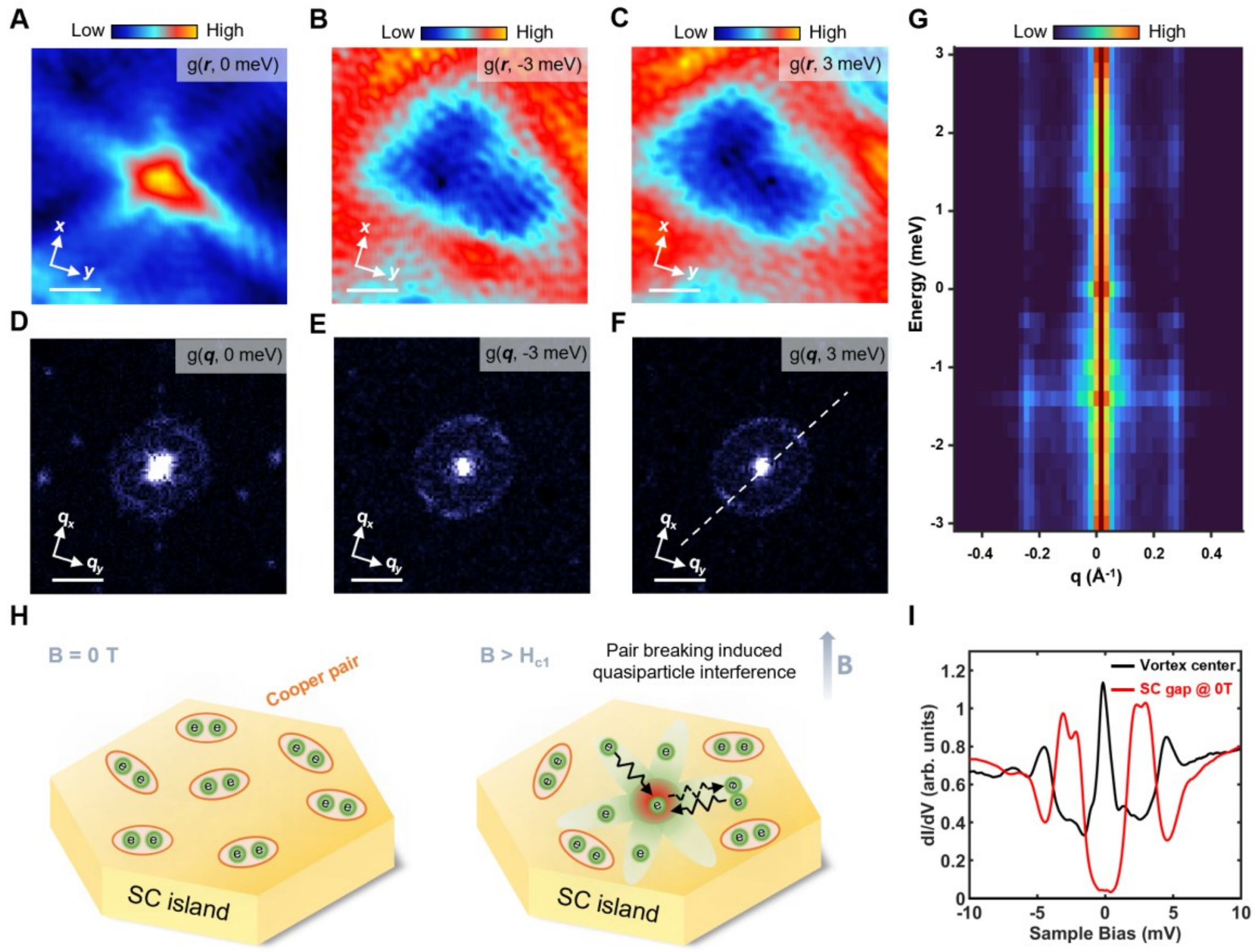

**Fig. 2| Anisotropic Abrikosov vortex and pair breaking induced quasiparticle interference.** **A,** Zero energy *dI/dV* mapping of magnetic vortex (50 nm×50 nm, $V_{bias}$ = 10 mV, $I_{Tunnel}$ = 500 pA, $V_{mod}$ = 0.1 mV, and sample bias when measured $V_{sample}$ = 0 mV, 1.0 T). **B-C,** *dI/dV* mapping images of same area in (**A**) at different energy, -3 meV for (**B**) and +3 meV for (**C**) (50 nm×50 nm, $V_{bias}$ = 10 mV, $I_{Tunnel}$ = 500 pA, $V_{mod}$ = 0.1 mV). **D-F**, Corresponded FFT pattern of 0 meV(**D**) -3 meV (**E**) and +3 meV (**F**). **G**, Summary of the FFT line profiles taken along write dash line shown in (**F**) at various energies, with same magnitude of scattering vector. **H**, Schematic of magnetic field induced pair breaking and quasiparticle interference. **I**, Typical *dI/dV* spectra taken on superconducting 1QL-$Bi_2Te_3$ surface at different position under different magnetic field, red line for 0T at clean surface and black line for 1.0T at vortex core center ($V_{bias}$ = 10 mV, $I_{Tunnel}$ = 500 pA, $V_{mod}$ = 0.1 mV). Scale bar: 10 nm (**A-C**); 0.3 Å$^{-1}$ (**D-F**).

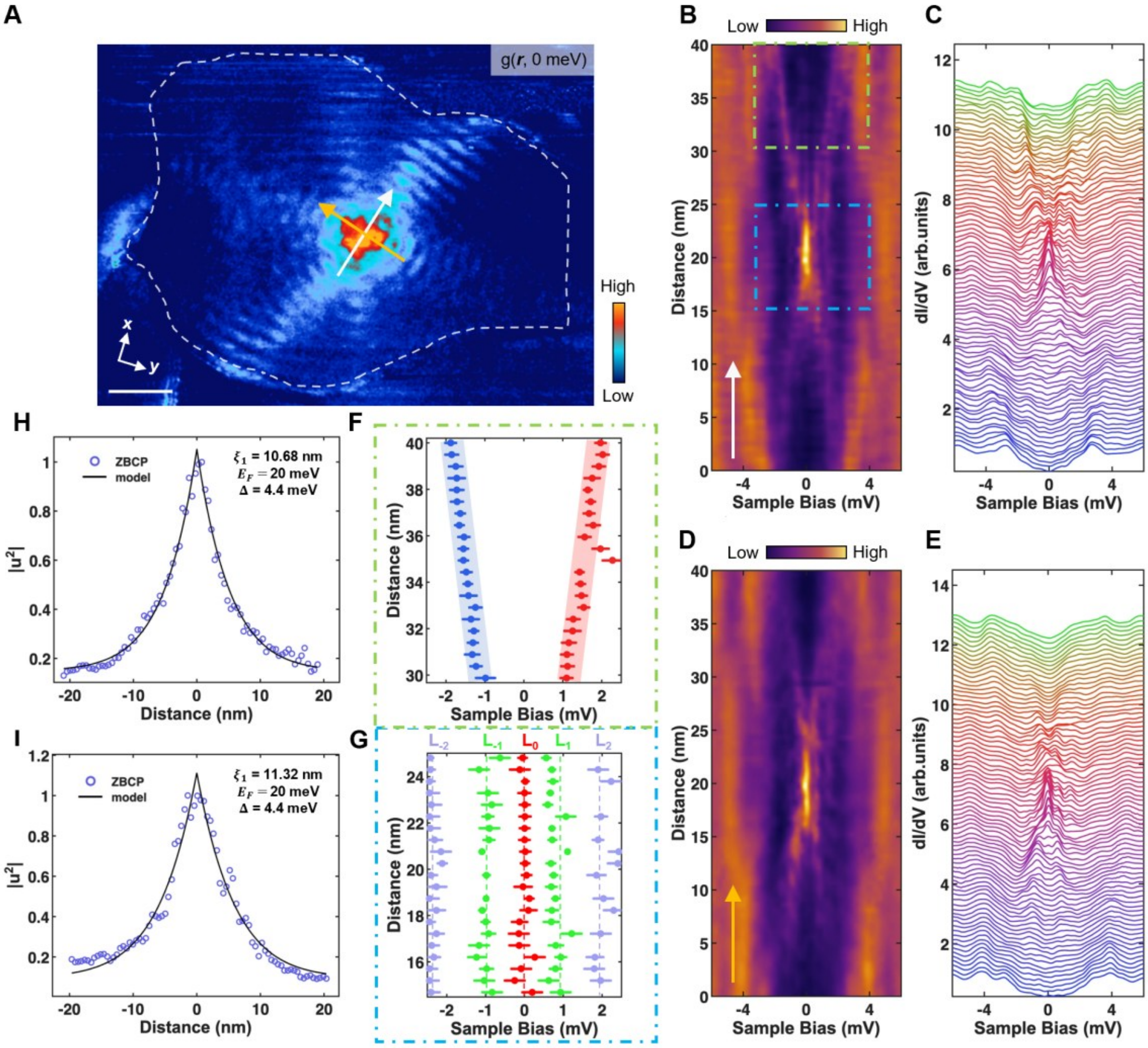

**Fig. 3| Single vortex in superconducting $Bi_2Te_3$ island and analysis of the vortex bound states. A,** Single vortex measured in superconducting $Bi_2Te_3$ island (100 nm×100 nm, $V_{bias}$ = 10 mV, $I_{Tunnel}$ = 500 pA, $V_{mod}$ = 0.1 mV, and sample bias when measured $V_{sample}$ = 0 mV), with clear sixfold symmetry. The boundary of 1QL-$Bi_2Te_3$ island is indicated by the white dashed line. **B-E,** Color plot and waterfall plot of *dI/dV* spectrums along different direction ($V_{bias}$ = 10 mV, $I_{Tunnel}$ = 1 nA, $V_{mod}$ = 0.1 mV), white arrow in (**A**) for **(B-C)**, and orange arrow for (**D-E**). **F,** Statistical analysis of the peak positions far away from vortex core center (10~20nm away from center; green dash rectangle in **B**). Double peaks are extracted and show linear dispersion along distance (bule and red dash line). **G,** Statistical analysis of the peak positions in vortex core (blue dash rectangle in **B**). Five discrete integer energy bound states are extracted, with no obviously dispersion. **H,** Comparison of measured ZBCP peak intensity and theoretical calculation of MZM spatial profile, the data are extracted from **(C)**, which is along the white arrow shown in **(A)**. The value of $\xi_0$ is fitted by the measured data, while the $\Delta$ and $E_F$ is given by experimental results. **I,** Comparison of measured ZBCP peak intensity and theoretical calculation of MZM spatial profile, the data are extracted from **(E)**, which is along the orange arrow shown in **(A).** Scale bar: 20 nm (**A**).

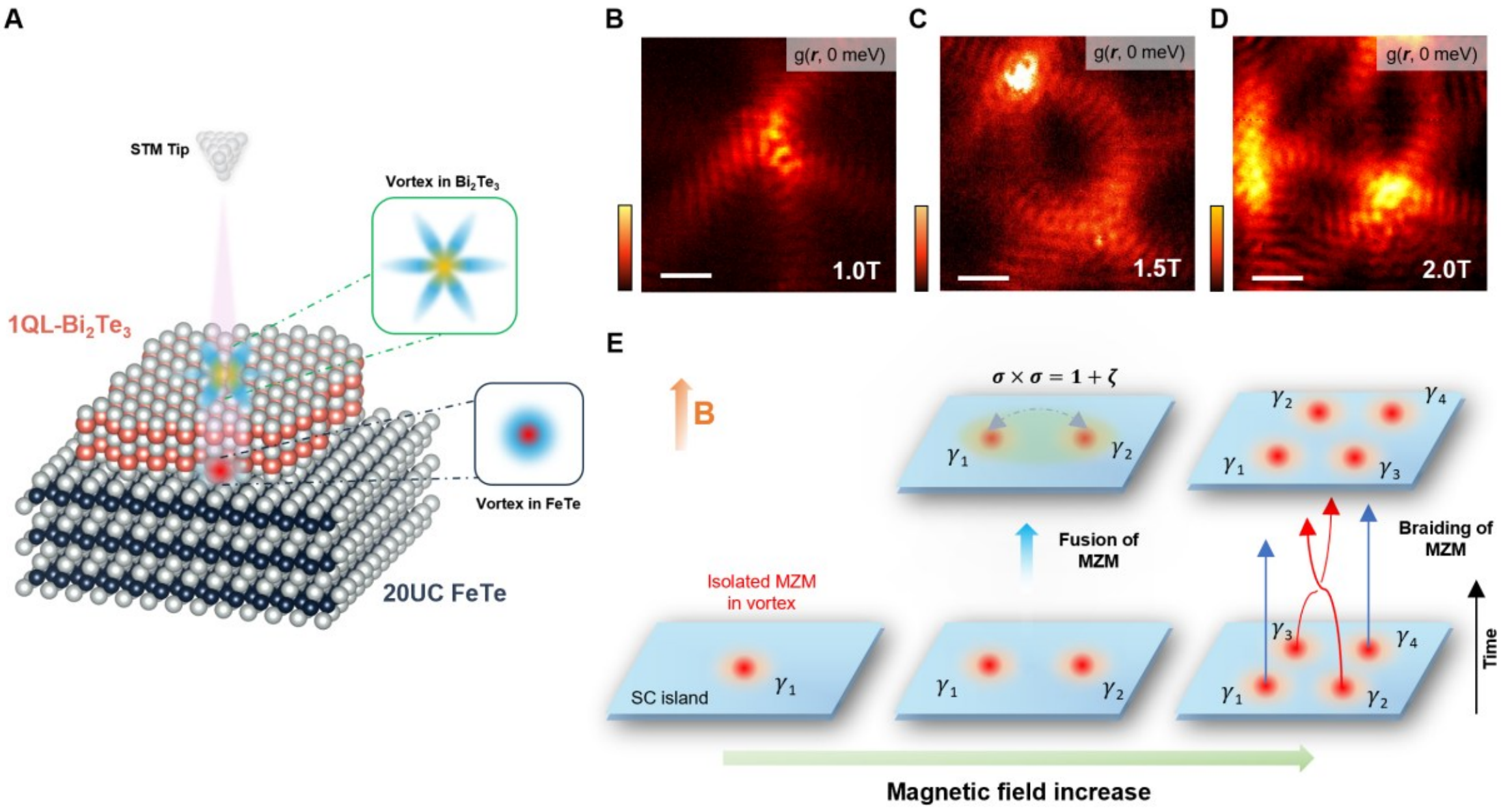

**Fig. 4| Schematic of hybrid vortices measurement and the concept of operation for MZM in superconducting island. A,** Schematic of simultaneously measured two vortices originating from different layers by STM tip, six-star shape vortex from $Bi_2Te_3$ and circle shape vortex from superconducting FeTe. **B-D**, Abrikosov vortices of 1QL-$Bi_2Te_3$/FeTe heterojunction measured under different perpendicular magnetic field at zero energy (68 nm×68 nm, $V_{bias}$ = 10 mV, $I_{Tunnel}$ = 1 nA, $V_{mod}$ = 100 uV). **E**, Schematic of changes in the number of vortices with increasing magnetic field in finite superconducting island and a possibility to realize the operation of MZM in 1QL-$Bi_2Te_3$/FeTe heterojunction. Scale bar: 13.6 nm (**B-D**).

**Acknowledgements**

We thank Yao Zhang, Renjie Zhang, Qiangsheng Lu and Hailang Qin for valuable discussions, Yang Qiu and Wenyu Hu for assistance in performing HR-STEM experiments at the Pico Center from SUSTech Core Research Facilities.

**Funding**

H. D. acknowledges support from the National Natural Science Foundation of China (Grants No.12488101), the Quantum Science and Technology-National Science and Technology Major Project (No. 2021ZD0302700), the New Cornerstone Science Foundation (No. 23H010801236), and the Fundamental and Interdisciplinary Disciplines Breakthrough Plan of the Ministry of Education of China (No. JYB2025XDXM411). B.L. acknowledges support from the Ministry of Science and Technology of China (2023YFA1407400), the National Natural Science Foundation of China (12374063, 92565305), the Cultivation Project of Shanghai Research Center for Quantum Sciences (LZPY2024), and the AI for Science Seed Program of Shanghai Jiao Tong University (2025AI4S-QY07). G. W. acknowledges support from States Key Laboratory of Quantum Functional Materials. Q. H. acknowledges support from the startup funding from Xiamen University. X. Liu acknowledges the support from the Quantum Science and Technology-National Science and Technology Major Project (No.2021ZD0302700), the National Natural Science Foundation of China (No. 12074133), the Shanghai Science and Technology Innovation Action Plan (No. 24LZ1400800) and the Project supported by Cultivation Project of Shanghai Research Center for Quantum Sciences (No. LZPY2024). X.-H. Pan is supported by the National Natural Science Foundation of China (Grant No. 12504190) and Guangdong Basic and Applied Basic Research Foundation (Grant No. 2026A1515011510). D.M. acknowledges support from the China Postdoctoral Science Foundation (Grant No. GZC20252230). F.Y. acknowledges the support by the fellowship awards from the Research Grants Council of Hong Kong (Project No. CityU JRFS2627-1S12).

**Author contributions**

H. D., G. W. and B. L. conceived the experiment. X. H., X. Z. and M. L. performed the MBE growth. X. H., Z. L., X. L., and D. M. performed STM measurements and the data analysis with the assistance from Q. H., Z. X. and Q. G.. X. Z. and M. P. performed the ARPES measurements. X. P. and Xin Liu performed the theoretical analysis. X. H. and Q. H. wrote the manuscript with contributions from all the authors.

**Competing financial interests**

The authors declare no competing interests.

**Data availability**

Source data will be provided with this paper. More data supporting of this study are available from the corresponding author upon reasonable request.

**Code available**

The code used for STM analysis is available from the corresponding author upon reasonable request.